\documentclass{article}
\usepackage{amsfonts,amssymb, amsmath}
\usepackage{graphicx}

\newtheorem{prop}{Proposition}
\makeatletter
\newcommand\xleftrightarrow[2][]{\ext@arrow 0099{\longleftrightarrowfill@}{#1}{#2}}
\def\longleftrightarrowfill@{\arrowfill@\leftarrow\relbar\rightarrow}
\makeatother

\def\nn{\nonumber }
\def\bq{ \begin{equation}}
\def\eq{ \end{equation}}
\def\ben{ \begin{eqnarray}}
\def\en{ \end{eqnarray}}

\begin{document}


\title{On exact discretization of the cubic-quintic Duffing oscillator}
\author{A.V. Tsiganov \\
\it\small St.Petersburg State University, St.Petersburg, Russia\\
\it\small e--mail:  andrey.tsiganov@gmail.com}
\date{}
\maketitle

\begin{abstract}
Application of  intersection theory to construction of $n$-point finite-difference equations associated with  classical integrable systems is discussed.  As an example, we present a few exact discretizations of  one-dimensional cubic and quintic Duffing oscillators sharing form of Hamiltonian  and canonical  Poisson bracket up to the integer scaling  factor.

 \end{abstract}

\section{Introduction}
\setcounter{equation}{0}
A completely integrable system on   symplectic manifold $M$ with  form $\omega$ of dimension $2n$ is defined by  $n$ smooth functions $f_1, \ldots, f_n$ in involution
\[ \{f_i,f_j \} = 0\,,\qquad i,j=1,\ldots,n \]
with the independent differentials $df_i$ at each cotangent space $T_x^*(M)$, $x\in M$.

If $c \in \mathbb{R}^n$ is  a regular value of  $ f = (f_1, \dots, f_n)$, then the corresponding level $X = f^{-1}(c)$ is a smooth $n$-dimensional  Lagrangian submanifold of $M$. Geometrically this means that locally around the regular value $c$ the map  $ f: M \rightarrow \mathbb{R}^n$ collecting the integrals of motion is a Lagrangian fibration, i.e. it is locally trivial and the fibers are Lagrangian submanifolds.

Let us consider a finite-difference equation
\bq\label{fd-eq}
\mathcal Y\left(P_{k-\ell},\ldots,P_k,\ldots P_{k+m};k\right)=0\,,\qquad k,\ell,m\in\mathbb Z.
\eq
relating $\ell+m+1$  points $P_i$ of submanifold $X$.  Ordinary finite difference equations of this type can be  viewed as a dynamical $\ell+m+1$-point map, see  \cite{hiet16}.

Because all  points $P_i$ in (\ref{fd-eq}) belong to  the given Lagrangian submanifold   we may suppose that the corresponding map preserves functions  $f_i$ and symplectic form $\omega$ up to a scaling factor.  Finite-difference equation sharing integrals of motion with the continuous time system and the  symplectic structure is the so-called exact discretization of  integrable systems. Nowadays, refactorization in the Poisson-Lie groups is viewed as one of the most universal constructions of  finite difference equations (\ref{fd-eq}), see  discussion in \cite{bob98,d91,hiet16,kuz02,mos91,sur03} and references within.

The idea is to identify $\ell+m+1$ points $P_{k-\ell},\ldots,P_{k+m}$ in (\ref{fd-eq}) with  intersection points of  $X$ with auxiliary curve $Y$.   If $X$ and $Y$ are algebraic, then we can consider the standard equation for  their intersection divisor
\[
div(X\cdot Y)=0
\]
as  the finite-difference equation (\ref{fd-eq}) for the corresponding completely integrable system. Here $div(X\cdot Y)$ is the intersection divisor of two algebraic varieties and $=$ is a suitable equivalence relation \cite{eh16,ful84,grif04,kl05}.  Our main objective is to study properties of such  $\ell+m+1$-point finite-difference equations for different integrable systems \cite{ts17a,ts17b,ts17c,ts17d,ts18d}.
In this paper  we restrict ourselves by consideration of  cubic and quintic nonlinear Duffing oscillators  in order to clarify our view point on relations between the exact discretizations and the intersection divisors.

Thus, we consider integrable systems on   two-dimensional plane $M$ with a pair coordinates $q,p$ and symplectic form $\omega=dp\wedge dq$.  Because any smooth curve on the plane is a Lagrangian submanifold, we can directly apply classical intersection theory \cite{bak97,gr} to exact discretization of one-dimensional Hamiltonian systems with the algebraic Hamilton function. Below we consider Hamiltonians $H(q,p)$  associated with hyperelliptic curves $X$ on the projective plane defined by equation
\[
X:\qquad y^2=a_{2g+2}x^{2g+2}+a_{2g+1}x^{2g+1}+\cdots+a_1x+a_0\,,\qquad a_j\in\mathbb C\,,
\]
at  $g=1,2$ and various intersections of $X$ with the line, quadric and cubic on the plane defined by
\bq\label{y-curve}
Y:\qquad y=\mathcal P(x)\,,\qquad \mathcal P(x)=b_mx^m+\cdots+b_0\,,\qquad m=1,2,3.
\eq
From now on  $x$ and $y$ mean coordinates on the projective plane, whereas $q$ and $p$  are coordinates on the phase space $M$.

\section{Cubic oscillator}
\setcounter{equation}{0}
The Hamilton function
\bq\label{cub-ham}
H(q,p)=p^2-a_4q^4-a_3q^3-a_2q^2-a_1q
\eq
and canonical Poisson bracket $\{q,p\}=1$ determine Hamiltonian equations
\bq\label{cub-ham-eq}
\dot{q}=\dfrac{\partial H}{\partial p}=2p\,,\qquad \dot{p}=-\dfrac{\partial H}{\partial q}=4a_4q^3+3a_3q^2+2a_2q+a_1
\eq
and equation of motion
\bq\label{cub-eqm}
\ddot{q}=8a_4q^3+6a_3q^2+4a_2q+2a_1\,,
\eq
for the generalized oscillator with the cubic nonlinearity \cite{mook73}.

At $a_3=a_1=0$ this integrable system is called a cubic Duffing oscillator without forcing.
 Duffing oscillators have received remarkable attention in recent decades due to the variety of their engineering applications. For instance  magneto-elastic mechanical systems, large amplitude oscillations of centri\-fu\-gal governor systems, nonlinear vibration of beams, plates and fluid flow induced vibration, seismic waves before earthquake, ecology or cancer dynamics,  financial fluctuations and so on are modeled by the nonlinear Duffing equations.

In the numerical integration of nonlinear differential equations, discretization of the nonlinear terms poses extra ambiguity in reducing the differential equation to a discrete difference equation. For instance, in the framework of the standard-like discretization differential equation
\bq\label{duff-eq}
\ddot{q}+Aq+Bq^3=0
\eq
 can be transformed to the  finite-difference equation
\[
\dfrac{q_{n+1}-2q_n+q_{n-1}}{h^2}+Aq_n+B(q_{n+1}+q_{n-1})q_n^2=0\,,
\]
where $h$ is a  discrete time interval \cite{pots81,pots82}.  This equation may be reduced to the expression with the mapping function $F(q_n)$
\[
q_{n+1}-2q_n+q_{n-1}=F(q_n)\,,\qquad F(q_n)=\dfrac{-Aq_n-Bq_n^3}{h^{-2}+1/2Bq_n^2}
\]
 or to the area preserving map on the plane
 \[
 p_{n+1}=p_n+\phi(q_n)\,,\qquad q_{n+1}=q_n+p_{n+1}\,,
 \]
 where $\phi(q_n)$ is a rational control function \cite{map03a,map03b}. This integrable map admits the invariant integral
\[
\tilde{H}=p_{n+1}^2+Aq_nq_{n+1}+Bq_n^2q_{n+1}^2\,,
\]
see details in \cite{map03a,map03b,pots81,pots82,rt86,sur89,sur03},  but it is not exact discretization of the Duffing oscillator, i.e.
 trajectories of the discrete flow do not coincide with the trajectories of the continuous flow.

\subsection{Exact discretization and intersection divisors}
In order to get exact discretization  sharing integrals of motion with the continuous time system and the  Poisson bracket we can start with the well-known analytical solutions $q(t)$ of the  Duffing equation (\ref{duff-eq}), which are expressed via Jacobi elliptic functions.

Indeed, let us consider the equation (\ref{duff-eq}) with initial condition
\[
q(0)=\alpha\,,\qquad \dot q(0)=0\,.
\]
For  $B>0$ and $A>-\alpha^2B$ periodic solution is
\[
q(t)=\alpha\,\mbox{cn}\left(2(A+2\alpha^2B)^{1/2}t\,;m\right)\,,\qquad m=\dfrac{\alpha^2B}{A+2\alpha^2B}\,.
\]
For  $B>0$ and $-B\alpha^2<A<-2B\alpha^2$ periodic solution reads as
\[
q(t)=\alpha\,\mbox{dn}\left(2B^{1/2}t\,;m\right)\,,\qquad m=2\left(1+\dfrac{A}{2\alpha^2}\right)\,.
\]
For $B<0$ and $A>-2B\alpha^2$  periodic solution has the form
\[
q(t)=\alpha\,\mbox{sn}\left(2(A+\alpha^2B)^{1/2}t\,;m\right)\,,\qquad m=-\dfrac{\alpha^2B}{A+\alpha^2B}\,.
\]
Here $\mbox{cn}(z;m)$ and  $\mbox{sn}(z;m)$  are the Jacobi elliptic functions. Discussion of the non-periodic solution can be also found  in \cite{pots81,pots82}.

Following \cite{bob98} we can construct exact discretizations of the Duffing equation using these explicit solutions and well-known addition theorems for Jacobi elliptic functions,  for instance
 \[
 \mbox{sn}(X+Y)=\dfrac{\mbox{sn}X\,\mbox{cn}Y\,\mbox{dn}Y+\mbox{sn}Y\,\mbox{cn}X\,\mbox{dn}X}{1-m^2\,\mbox{sn}^2X\,\mbox{sn}^2Y}\,.
 \]
However, it is more easy and  convenient to apply standard algorithms of the intersection theory for this purpose.

In order to apply the intersection theory to the exact discretization of the cubic oscillator we put $H (q, p) =E$ and consider the corresponding level curve $X$ on the projective plane defined by equation
\bq\label{ell-curve}
X:\quad y^2=f(x)\,,\qquad f(x)=a_4x^4+a_3x^3+a_2x^2+a_1x+a_0\,.
\eq
where $E=a_0$.  Any partial solution $q(t)$ and $p(t)$ of the Hamiltonian equations  (\ref{cub-ham-eq}) at $t=t_i$ is a point  $P_i=(x_i,y_i)$ of $X$ with abscissa $x_i=q(t_i)$ and ordinate $y_i=p(t_i)$. It allows us to study relations between points of $X$ instead of relations between solutions of the differential equations (\ref{cub-ham-eq}).

Let $X$ be a smooth nonsingular algebraic  curve  on a projective plane.  Prime divisors are rational points on $X$  denoted $P_i = (x_i, y_i)$  and $P_\infty$ is a point at infinity. Divisor
\[D = \sum m_iP_i\,,\qquad m_i\in \mathbb Z\]
is a formal sum of prime divisors, and the degree of  divisor $D$ is a sum deg$D=\sum m_i$ of  multiplicities of points in  support of the divisor.  Group of divisors is an additive Abelian group under the formal addition rule
 \[\sum m_i P_i+\sum n_i P_i=\sum (m_i+n_i) P_i\,.\]
 Two divisors $D, D'\in \mbox{Div} X$ are linearly equivalent
\[D\approx D'\]
if their difference $D-D'$ is principal divisor
\[
D-D'=div(\psi)\equiv 0\quad \mathrm{mod\, Prin}X\,,
\]
i.e. divisor of  rational function $\psi$ on $X$.

Intersection divisor of $X$ with some auxiliary smooth nonsingular plane curve $Y$
\[
div(X\cdot Y)=0 \, \mathrm{mod\, Prin}X
\]
is equal to zero with respect to the linear equivalence of divisors. It allows us to identify  intersection divisor with some finite-difference equation (\ref{fd-eq})
\bq\label{fd-ed-d}
 \mathcal Y\left(P_{k-\ell},\ldots,P_k,\ldots, P_{k+m};k\right)=\sum_{i=k-\ell}^{k+m} m_iP_i+\sum_{i} n_iP^{(k)}_i= 0\, \mathrm{mod\, Prin}X\,.
\eq
Here we divide intersection divisor in two parts
\[
div(X\cdot Y)=\sum_{i=k-\ell}^{k+m} m_iP_i+\sum_{i} n_iP^{(k)}_i
\]
 where prime divisors $P^{(k)}_i$ are parameters of discretization implicitly depending on $k$.

\subsection{Examples of the intersections}
Let us consider the intersection of  plane curve $X$ (\ref{ell-curve}) with a parabola
\[
Y:\qquad y=\mathcal P(x)\,,\qquad \mathcal P(x)=b_2x^2+b_1x+b_0
\]
and the corresponding intersection divisor $div(X\cdot Y)$ of degree four, see \cite{bak97}, p.113 or  \cite{gr}, p.166. Following Abel we substitute $y=\mathcal P(x)$ into (\ref{ell-curve}) and obtain the  so-called Abel polynomial
\[
\psi(x)=\mathcal P(x)^2-f(x)\,.
\]
Divisor of this polynomial on $X$ coincides with $div(X\cdot Y)$, i.e. roots of this polynomial are abscissas of  intersection points $P_1,P_2,P_3$ and $P_4$ forming support of the intersection divisor $div(X\cdot Y)$.

At $b_2=\sqrt{a_4}$ one of the intersection points is $P_\infty$, see examples in Figure 1.
\begin{figure}[!ht]
\begin{minipage}[h]{0.49\linewidth}
\center{\includegraphics[width=0.85\linewidth, height=0.2\textheight]{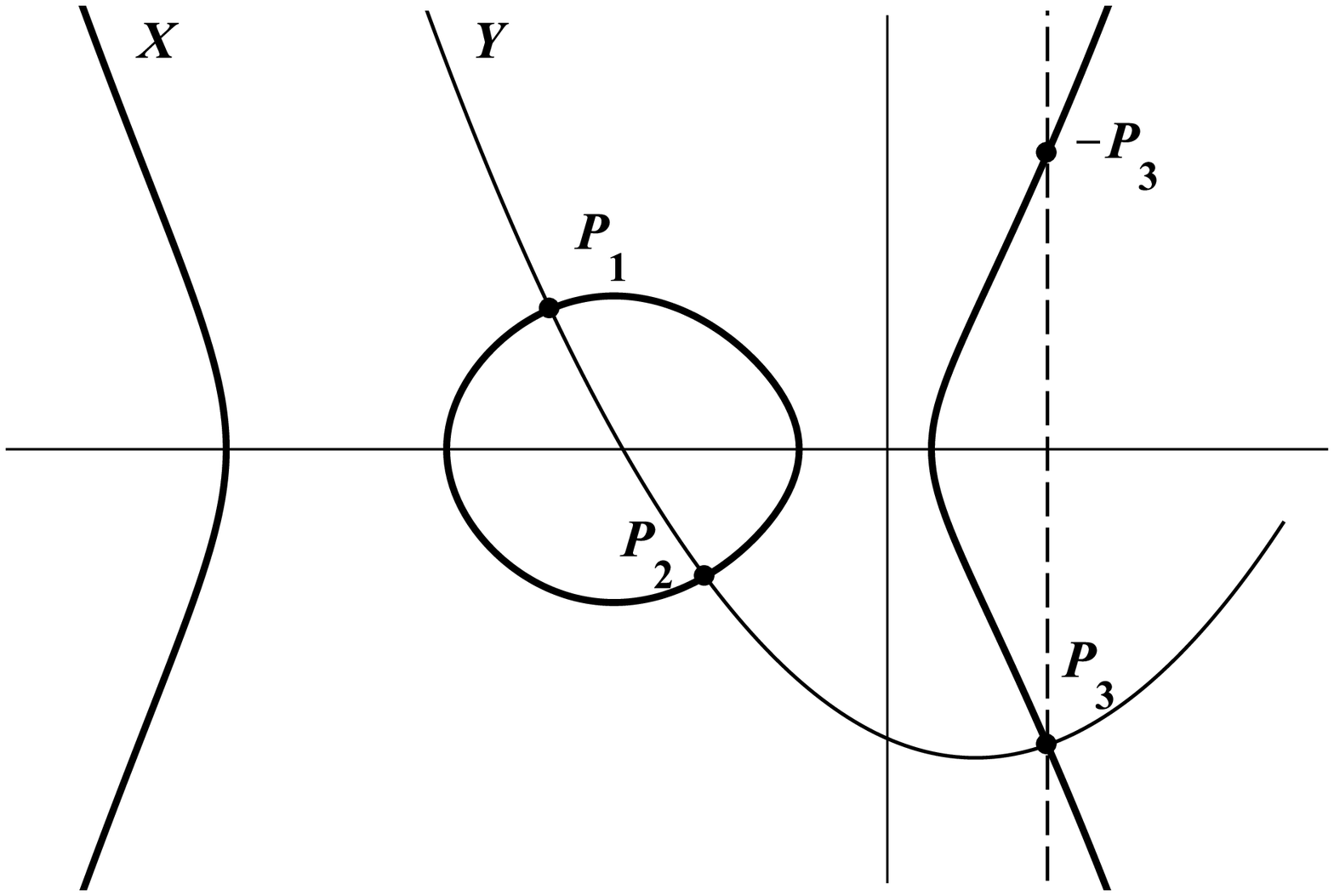} \\ a)  $(P_1+P_2)+P_3+P_\infty=0$}
\end{minipage}
\hfill
\begin{minipage}[h]{0.49\linewidth}
\center{\includegraphics[width=0.85\linewidth,height=0.2\textheight]{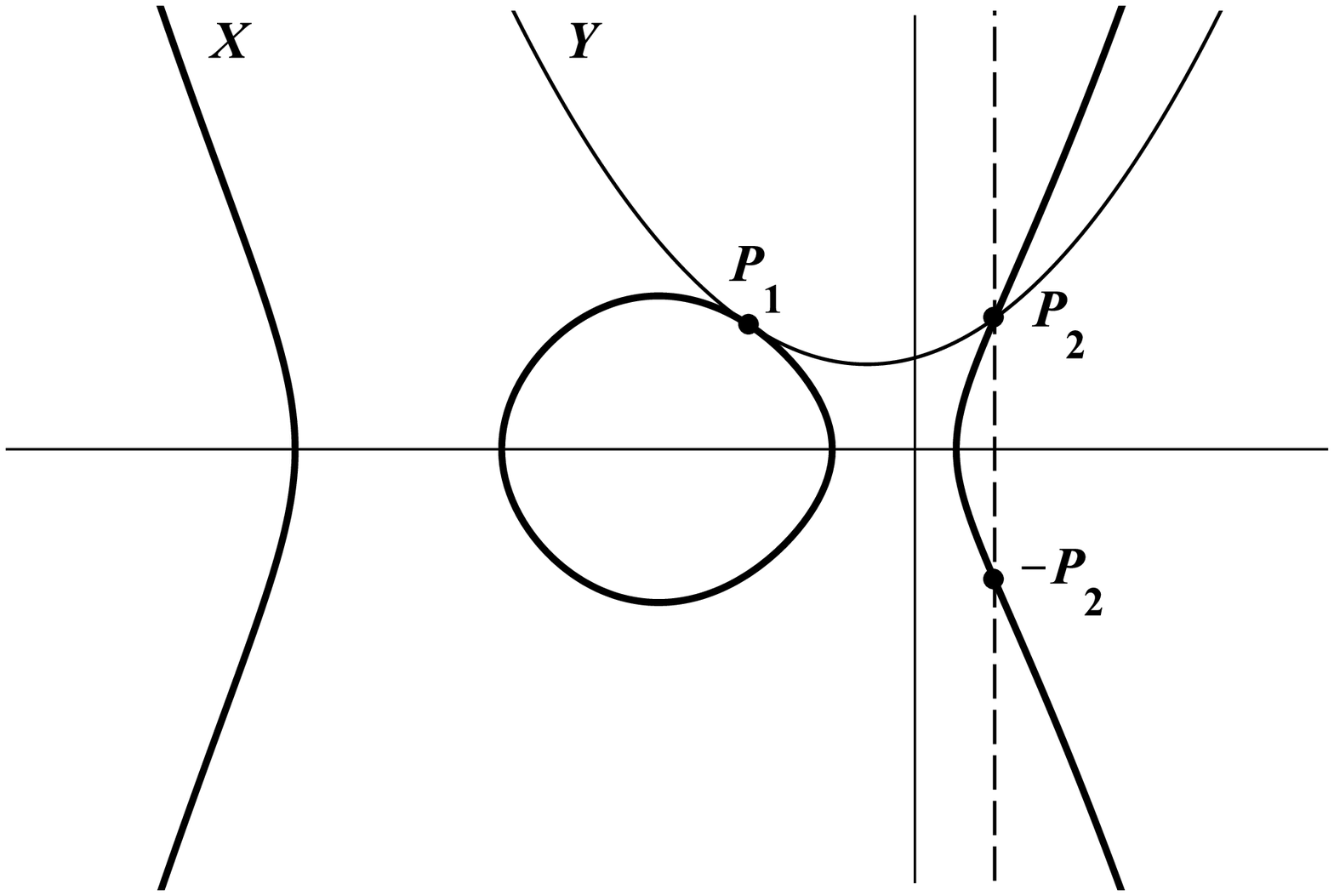} \\  b) $(2P_1)+P_2+P_\infty=0$}
\end{minipage}
\caption{Interection of  curve $X$  (\ref{ell-curve}) with parabola $Y:\,y=\sqrt{a_4}x^2+b_1x+b_0$}
\end{figure}
\par\noindent
In this case polynomial $\psi(x)$ is equal to
\[\begin{array}{rcl}
\psi(x)&=&(2b_1b_2-a_3)x^3+(2b_0b_2+b_1^2-a_2)x^2+(2b_0b_1-a_1)x+b_0^2-a_0\nn\\
\nn\\
&=&(2b_1b_2-a_3)(x-x_1)(x-x_2)(x-x_3)\,.
\end{array}
\]
Equating coefficients of $\psi$  one gets relation between abscissas of  the remaining rational points $P_1,P_2$ and $P_3$ in support of the intersection divisor
\bq\label{add-gen}
x_1+x_2+x_3=-\dfrac{2b_0b_2+b_1^2-a_2}{2b_1b_2-a_3}\,.
\eq
If $P_i\neq P_j$ as in Figure 1a, we can define parabola $Y$  using the Lagrange interpolation by any pair of points  $ (P_1, P_2) $, $ (P_1, P_3) $ or $ (P_2, P_3) $.
For instance, taking the following pair of points $(P_1,P_2)$ one gets
\[
\mathcal P(x)=b_2x^2+b_1x+b_0=\sqrt{a_4}(x-x_1)(x-x_2)+\dfrac{(x-x_2)y_1}{x_1-x_2}+\dfrac{(x-x_1)y_2}{x_2-x_1},
\]
which allows us to determine  $b_2,b_1,b_0$ as functions on $x_{1,2}$ and $y_{1,2}$. Substituting coefficients of $\mathcal P (x) $ into the equation (\ref{add-gen}) we obtain an explicit expression for abscissa  $x_3$  as a function of coordinates  $x_ {1,2} $ and $y_ {1,2} $
\bq\label{add-ell}
x_3=-x_1-x_2+\phi(x_1,y_1,x_2,y_2)\,,\qquad \phi=-\dfrac{2b_0b_2+b_1^2-a_2}{2b_1b_2-a_3}
\eq
If we have a double intersection point, for instance $P_1=P_3$ as in Figure 1b, then
\bq\label{doub-ell}
x_2=-2x_1+\phi(x_1,y_1)\,,\qquad \phi=-\dfrac{2b_0b_2+b_1^2-a_2}{2b_1b_2-a_3}\,,
\eq
where function $\phi(x_1,y_1)$ is defined by $\mathcal P(x)$ due to the  Hermite interpolation
\[
\mathcal P(x)=b_2x^2+b_1x+b_0=\sqrt{a_4}(x-x_1)^2+\dfrac{(x-x_1)(4a_4x_1^3+3a_3x_1^2+2a_2x_1+a_1)}{2y_1}+y_1\,.
\]

In modern terms, we consider two partitions of the intersection divisor
\[
div(X\cdot Y)=(P_1+P_2)+P_3+P_\infty\qquad\mbox{and}\qquad
div(X\cdot Y)=(2P_1)+P_2+P_\infty.
\]
Using brackets $(.)$  we separate a part of  the intersection divisor which is necessary for polynomial interpolation of auxiliary curve $Y$. Because
\[
div(X\cdot Y)=0\,,
\]
these partitions can be rewritten as addition and doubling of prime divisors
\[
-P_3=P_1+P_2\,,\qquad -P_2=2P_1\,,
\]
 where we use standard hyperelliptic  inversion  $(x,y)\to (x,-y)$, see Figure 1.

At $b_2\neq \sqrt{a_4}$ support of the intersection divisor consists of four rational points $P_i\neq P_\infty$ up to multiplicity.
Let us consider the following partitions of this divisor
\[
div(X\cdot Y)=(P_1+P_2+P_3)+P_4\,,\quad div(X\cdot Y)=(2P_1+P_2)+P_3\,,\quad
div(X\cdot Y)=(3P_1)+P_2\,,
\]
see Figure 2. In the first case parabola $Y$ is defined by the Lagrange interpolation  using three ordinary points $P_1, P_2$ and $P_3$. In the second and third cases  parabola $Y$ is defined by the Hermite interpolation using either double  and ordinary  points $2P_1, P_2$  or one triple point $3P_1$, respectively.
\begin{figure}[!ht]
\begin{minipage}[h]{0.32\linewidth}
\center{\includegraphics[width=0.85\linewidth, height=0.2\textheight]{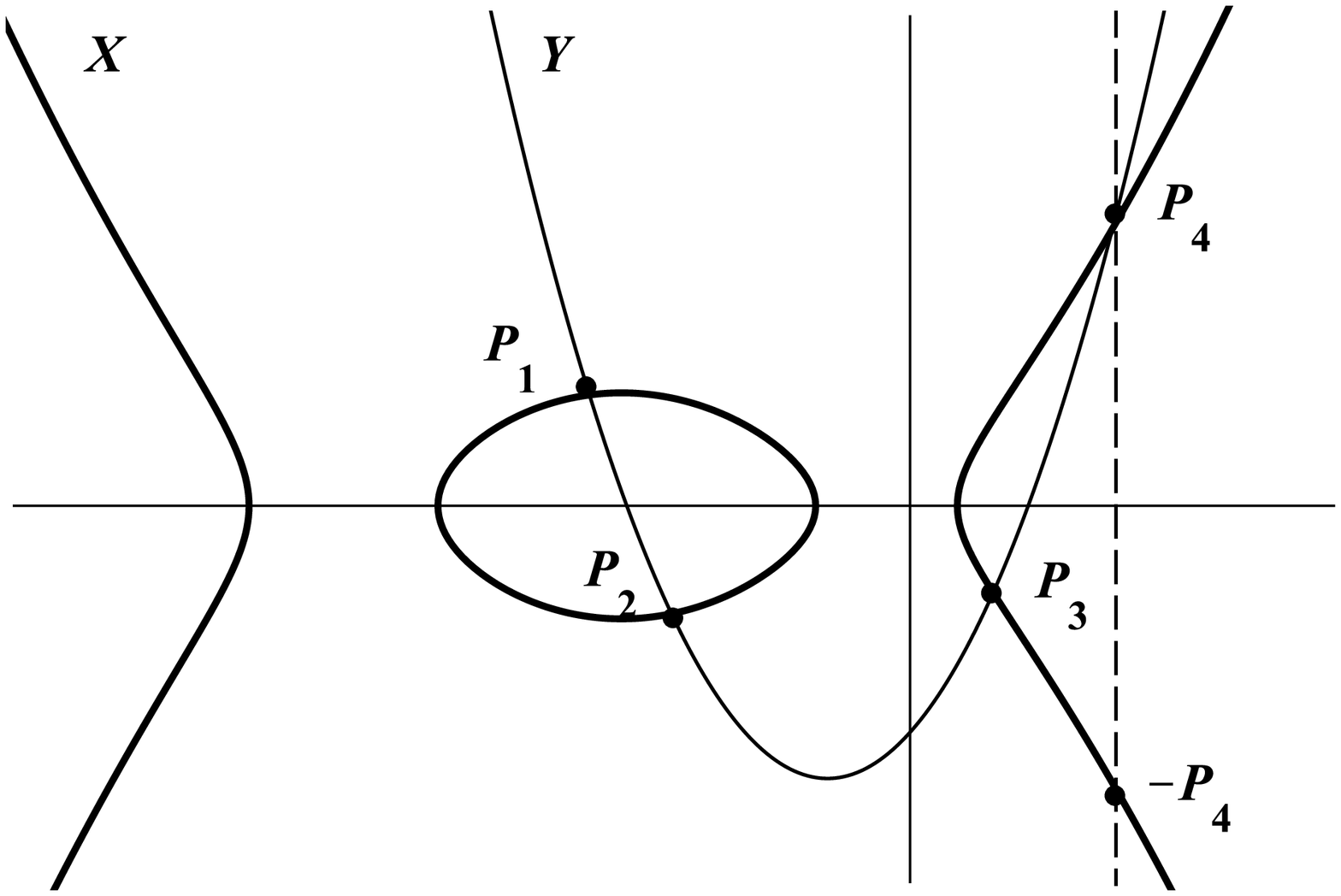} \\  a) $(P_1+P_2+P_3)+P_4=0$}
\end{minipage}
\hfill
\begin{minipage}[h]{0.32\linewidth}
\center{\includegraphics[width=0.85\linewidth,height=0.2\textheight]{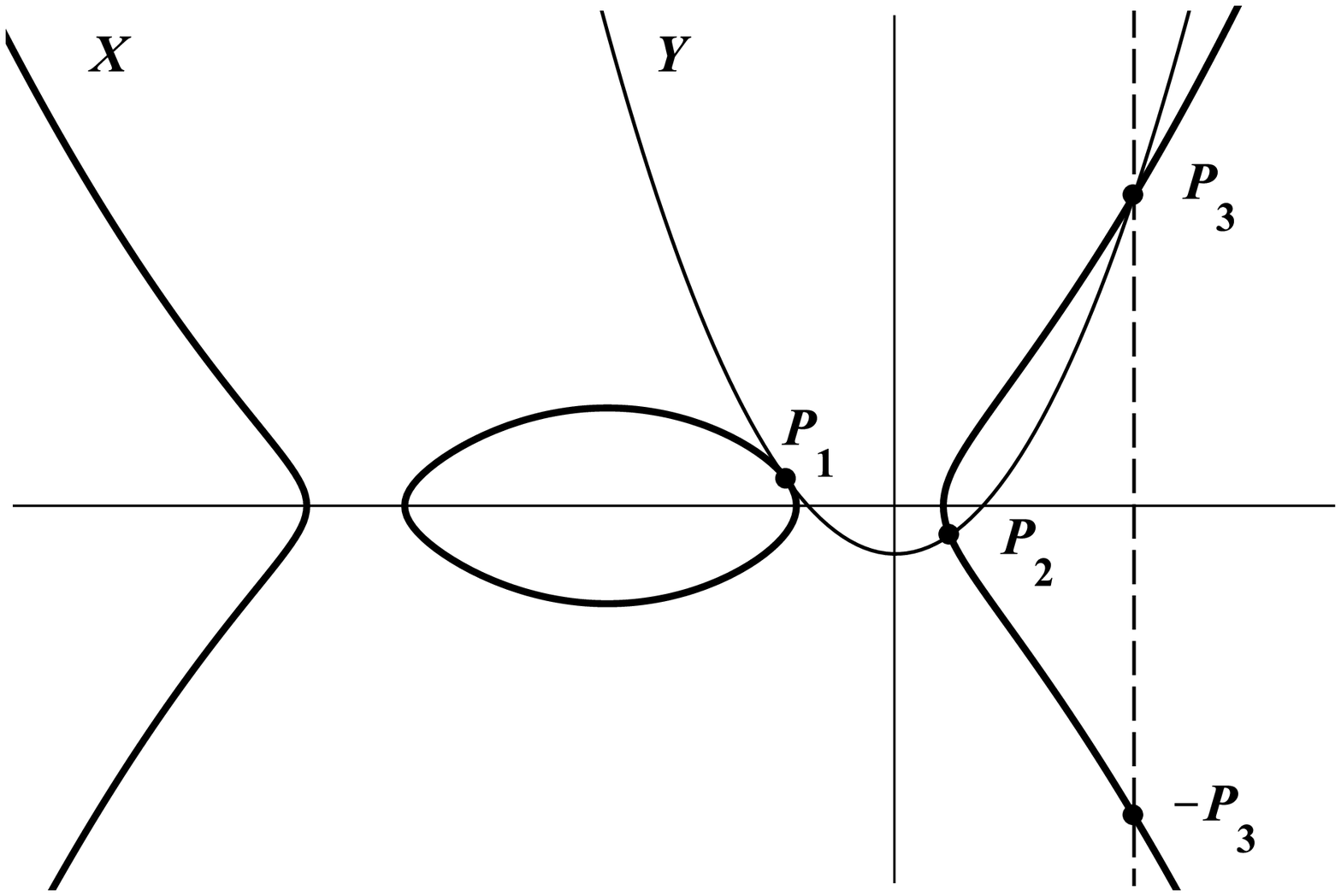} \\  b) $(2P_1+P_2)+P_3=0$}
\end{minipage}
\hfill
\begin{minipage}[h]{0.32\linewidth}
\center{\includegraphics[width=0.85\linewidth,height=0.2\textheight]{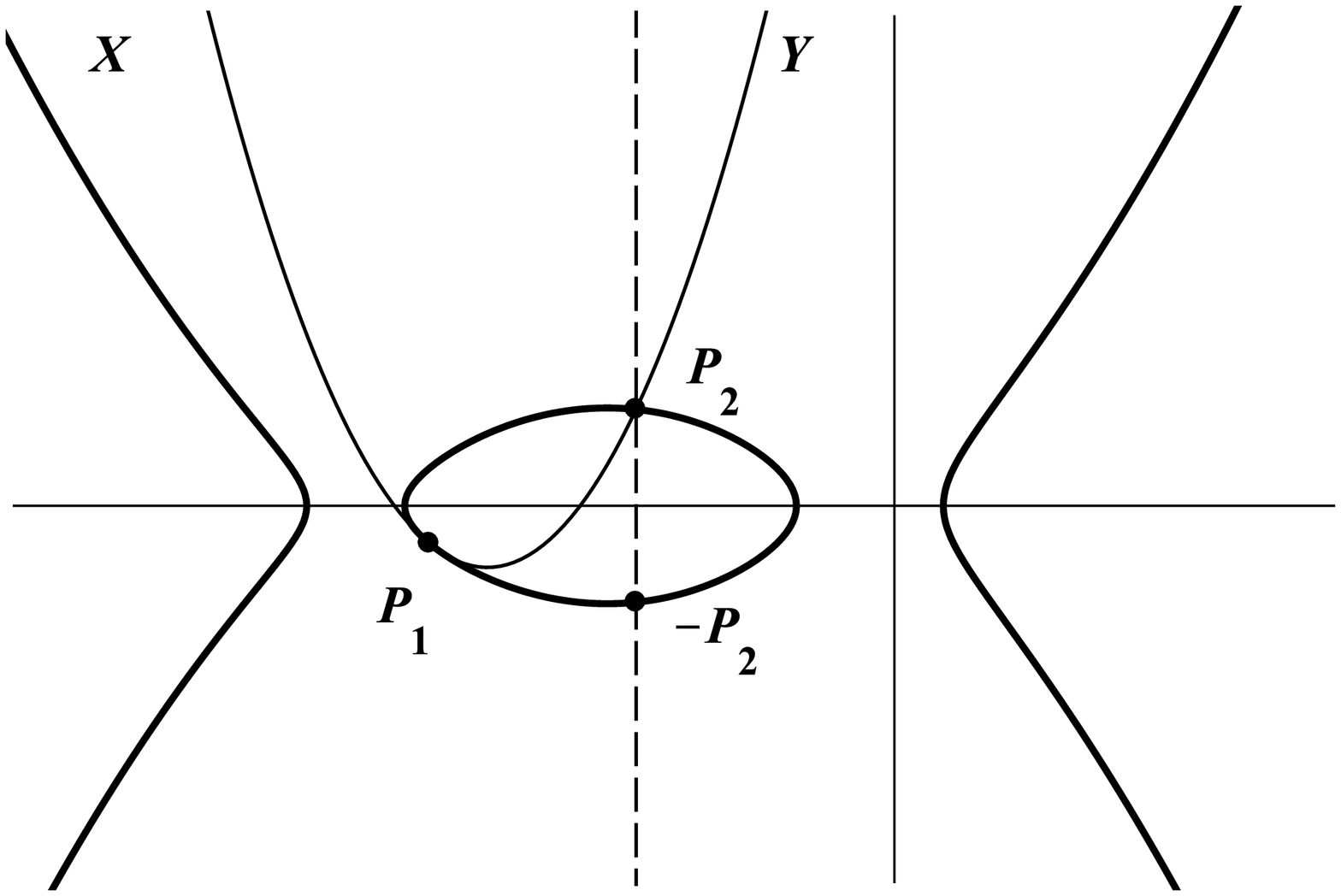} \\  c) $(3P_1)+P_2=0$}
\end{minipage}
\caption{Intersection  of  curve $X$  (\ref{ell-curve}) with parabola $Y:\,y=b_2x^2+b_1x+b_0$}
\end{figure}

In the first case abscissa of the fourth intersection point is
\bq\label{add-ell4}
x_4=-x_1-x_2-x_3+\varphi(x_1,x_2,x_3,y_1,y_2,y_3)\,,\qquad\varphi=-\dfrac{a_3-2b_1b_2}{a_4-b_2^2}\,,
\eq
where function $\varphi$  is defined using coefficients of  quadratic polynomial  $\mathcal P(x)=b_2x^2+b_1x+b_0$
\bq\label{lag-cub}
\mathcal P(x)=\dfrac{(x-x_2)(x-x_3)y_1}{(x_1-x_2)(x_1-x_3)}
+\dfrac{(x-x_1)(x-x_3)y_2}{(x_2-x_1)(x_2-x_3)}+\dfrac{(x-x_1)(x-x_2)y_3}{(x_3-x_1)(x_3-x_2)}\,.
\eq

In the second case expression for the abscissa looks like
\bq\label{add-21}
x_3=-2x_1-x_2+\varphi(x_1,x_2,y_1,y_2)\,,\qquad \varphi=-\dfrac{a_3-2b_1b_2}{a_4-b_2^2}\,.
\eq
Here function  $\varphi$ is defined via  coefficients of the same polynomial  $\mathcal P(x)=b_2x^2+b_1x+b_0$
and Hermite interpolation formulae
\[
\mathcal P(x)=\frac{(x-x_1)^2y_2-(x-2x_1+x_2)(x-x_2)y_1}{(x_1-x_2)^2}+\dfrac{(x-x_1)(x-x_2)(4a_4x_1^3+3a_3x_1^2+2a_2x_1+a_1)}{2y_1(x_1-x_2)}
\]

In the third case, when we consider tripling  the prime divisor on  $X$
\[
(x_2,y_2)=3(x_1,y_1)\,,
\]
second abscissa is equal to
\bq\label{trip-gen}
x_2= -3x_1+\varphi(x_1,y_1)\,,\qquad\varphi=-\dfrac{a_3-2b_1b_2}{a_4-b_2^2}\,,
\eq
where function $\varphi$ is defined  via  coefficients of the polynomial
\bq\label{trip-pol}
\begin{array}{rcl}
\mathcal P(x)=b_2x^2+b_1x+b_0&=&-\dfrac{(x-x_1)^2 (4 a_4 x_1^3+3 a_3 x_1^2+2 a_2 x_1+a_1)^2}{8 y_1^3}\\
\\
&+&\dfrac{(x-x_1)\Bigl(x \bigl(6 a_4 x_1^2+3 a_3 x_1+a_2\bigr)-2 a_4 x_1^3+a_2 x_1+a_1\Bigr)}{2 y_1}+y_1\,.
\end{array}
\eq

At $b_2=0$ we have the intersection divisor of $X$ with line $Y$, which can be represented in the following form
\[
div(X\cdot Y)=(P_1+P_2)+P_3+P_4\,.
\]
It means that line $Y$ is interpolated by two points $P_1$ and $P_2$
\[
\mathcal P(x)=b_1x+b_0=\dfrac{x-x_2}{x_1-x_2}y_1+\dfrac{x-x_1}{x_2-x_1}y_2\,,
\]
whereas abscissas of remaining two points $P_3$ and $P_4$ are the roots of polynomial
\[
\dfrac{\psi(x)}{(x-x_1)(x-x_2)}=a_4x^2+\bigl(a_4(x_1+x_2)+a_3\bigr)x+a_4(x_1^2+x_1x_2+x_2^2)+a_3(x_1+x_2)+a_2-b_1^2\,.
\]
Thus,  $x_{3,4}$ are algebraic functions on  coordinates $x_{1,2}$ and $y_{1,2}$
\bq\label{line-34}
x_{3,4}=-\dfrac{x_1}{2}-\dfrac{x_2}{2}-\dfrac{a_3}{2a_4}\pm\dfrac{\sqrt{\alpha(x_1,x_2,y_1,y_2)}}{2a_4}
\eq
where
\bq\label{alpha}
\alpha(x_1,x_2,y_1,y_2)=4a_4\left(\left(\dfrac{y_1-y_2}{x_1-x_2}\right)^2-a_2\right)+a_3^2-2a_4a_3(x_1+x_2)-a_4^2(3x_1^2+2x_1x_2+3x_2^2)\,.
\eq

In the generic cases, using intersection divisors of plane curve $X$ with  auxiliary curves
\[
Y:\qquad y=b_Nx^N+b_{N-1}x^{N-1}+\cdots+b_0\,,\qquad N=1,2,3,\ldots
\]
we can describe multiplication of the prime divisor on any integer $P_1=nP_2$, which is a key ingredient of the modern elliptic curve cryptography, and other configurations of the prime divisors entering into the intersection divisor.

All the relations between abscissas $x_k$ (\ref{add-gen}-\ref{line-34}) are well known, here we only repeat the fairly simple calculations based on Abel's ideas and their geometric interpretation proposed by Clebsch, see the historical comments in \cite{kl05}.  The modern intersection theory  gives a common language for the compact description of these partial cases of intersections  \cite{eh16,ful84}, whereas modern cryptography  equips us with the effective algorithms for such computations \cite{hand06}.

\subsection{Examples of finite-difference equations}
Our aim is to interpret well-studied  relations between prime divisors as the finite-difference equations (\ref{fd-eq}) realizing various exact discretizations of the given  Hamiltonian system and to study the properties of the corresponding discrete maps. For this purpose, we will identify partial  solutions $q (t) $ and $p (t) $ of the Hamilton equations (\ref{cub-ham-eq})
\[
\dot{q}=2p\,,\qquad \dot{p}=4a_4q^3+3a_3q^2+2a_2q+a_1
\]
at  $t=t_i$ with a prime divisor $P_i=(x_i,y_i)$, where $x_i=q(t_i)$ and $y_i=p(t_i)$.

For instance, substituting
\[x_1=q_1,\quad y_1=p_1\,,\qquad x_2=q_{2}\,,\quad y_2=p_{2}\,,\qquad
x_3=q_{3}\,,\quad y_3=p_{3}
\]
in  (\ref{add-gen}) and  $y_3=-\mathcal P(x_3)$ one gets finite-difference equations
\bq\label{add-map-1}
\begin{array}{rcl}
q_1+q_2+q_3&=&\phi(q_1,p_1,q_2,p_2)\,,\\
\\
\sqrt{a_4}(q_1-q_2)(q_2-q_3)(q_3-q_2)&=&q_3(p_1-p_2)+p_3(q_1-q_2)+q_1p_2-q_2p_1\,,
\end{array}
\eq
where $\phi=\phi_1/\phi_2$ is the  rational function on variables $q_1,p_1$ and $q_2,p_2$
\[\begin{array}{rcl}
\phi_1&=&a_2-a_4(q_1^2+4q_1q_2+q_2^2)
+\dfrac{2\sqrt{a_4}\bigl(q_1p_1-2(q_1p_2-q_2p_1)-q_2p_2\bigr)}{q_1-q_2}
-\dfrac{(p_1-p_2)^2}{(q_1-q_2)^2}
\\
\\
\phi_2&=&2\sqrt{a_4}\left(\dfrac{p_1-p_2}{q_1-q_2}-\sqrt{a_4}(q_1+q_2)\right)-a_3\,.
\\
\end{array}
\]
We can directly verify the following properties of the corresponding discrete mapping.
\begin{prop}
Relations  (\ref{add-map-1} ) determine 3-point mapping $M\times M\to M$
\[
\left(
  \begin{array}{c}
    q_{1},q_{2}\\
    p_{1},p_{2} \\
  \end{array}\right)
\xrightarrow[]{}\left(
  \begin{array}{c}
    q_{3} \\
    p_{3} \\
  \end{array}
\right)\,,
\]
preserving the form of Hamiltonain (\ref{cub-ham}) and  Poisson bracket, i.e.  from
$\{q_1,p_1\}=1$, $\{q_2,p_2\}=1$ and  (\ref{add-map-1}) will follow that $\{q_3,p_3\}=1$\,.
\end{prop}

In order to get an iterative system of finite-difference equations we identify a part of abscissas of the intersection points with
arbitrary numbers
\[
x_i=\lambda_{ik}\,,\qquad y_i=\mu_{ik}=\pm \sqrt{f(\lambda_{ik})}\,,\qquad \lambda_{ik}\in\mathbb C\,.
\]
In this case finite-difference equations (\ref{fd-ed-d}) implicitly depend on the independent variable
$k$ via parameters of discretization $\lambda_{ik}$.  For instance, addition of prime divisors
 \[
P_3=P_1+P_2\,,
\]
at
\[x_1=q_k,\qquad y_1=p_k\,,\quad x_3=q_{k+1}\,,\quad y_3=p_{k+1}
\qquad\mbox{and}\qquad x_2=\lambda_k\,,\qquad y_2=\mu_k\]
determines the following  iterative system of 2-point invertible mappings
\bq\label{add-map}
q_{k+1}=-q_k-\lambda_k+\phi(q_k,\lambda_k)\,,\qquad
p_{k+1}=-(b_2q_{k+1}^2-b_1q_{k+1}-b_0)\,,
\eq
where $\phi$ is given by (\ref{add-ell}) and
\[b_2=\sqrt{a_4}\,,\qquad
b_1=-\sqrt{a_4}(q_k+\lambda_k)+\dfrac{q_k-\mu_k}{q_k-\lambda_k}\,,\qquad
b_0=\sqrt{a_4}q_k\lambda_k+\dfrac{q_k\mu_k -\lambda_kp_k}{q_k-\lambda_k}\,.
\]
Here $\lambda_k$ are arbitrary numbers, whereas the corresponding ordinates
\[
\mu_k=\pm\sqrt{a_4\lambda_k^4+a_3\lambda_k^3+a_2\lambda_k^2+a_1\lambda_k+H}\,,\qquad
H=p_k^2-a_4q_k^4-a_3q_k^3-a_2q_k^2-a_1q_k
\]
are the functions on the phase space $M$. We have to use this fact to calculate Poisson bracket between variables  $q_{k+1}$ and $p_{k+1}$ (\ref{add-map}), obtained from  variables $q_{k}$ and $p_{k}$.

\begin{prop}
Relations (\ref{add-map}) determine  iterative system of 2-point invertible mappings
\[
\cdots\,\xrightarrow[\lambda_{k-2}]{}\left(
  \begin{array}{c}
    q_{k-1} \\
    p_{k-1} \\
  \end{array}\right)\xrightarrow[\lambda_{k-1}]{}\left(
  \begin{array}{c}
    q_{k} \\
    p_{k} \\
  \end{array}
\right)
\xrightarrow[\lambda_{k}]{}\left(
  \begin{array}{c}
    q_{k+1} \\
    p_{k+1} \\
  \end{array}
\right)\xrightarrow[\lambda_{k+1}]{}\,\cdots
\]
\vskip0.1truecm
\par\noindent preserving the form of Hamilton function
\bq\label{ell-dham}\begin{array}{rcl}
H&=&p_{k-1}^2-a_4q_{k-1}^4-a_3q_{k-2}^3-a_2q_{k-1}^2-a_1q_{k-1}\\ \\
&=&p_{k\phantom{+1}}^2-a_4q_{k\phantom{+1}}^4-a_3q_{k\phantom{+1}}^3-a_2q_{k\phantom{+1}}^2-a_1q_{k\phantom{+1}}\\ \\
&=&p_{k+1}^2-a_4q_{k+1}^4-a_3q_{k+2}^3-a_2q_{k+1}^2-a_1q_{k+1}\cdots\\
\end{array}
\eq
and Poisson bracket, i.e. from $\{q_k,p_k\}=1$ and (\ref{add-map}) will follow that $\{q_{k+1},p_{k+1}\}=1$.
\end{prop}
The proof is a straightforward calculation.

Substituting
\[x_1=q_k,\qquad y_1=p_k\,,\qquad x_2=q_{k+1}\,,\qquad y_2=p_{k+1}\]
in (\ref{doub-ell}) and  (\ref{trip-gen}) one gets two other iterative systems of 2-point mappings
\[
q_{k+1}=-q_k+\phi(q_k)\,,\qquad p_{k+1}=-(b_2q_{k+1}^2-b_1q_{k+1}-b_0)\,,
\]
associated with multiplication of prime divisor on integer $(x_2,y_2)=N(x_1,y_1)$ at $N=2,3$.
For the cubic Duffing oscillator at $a_3=a_1=0$ we present these mapping explicitly
\bq\label{M-map}
\begin{array}{ll}
N=2\,,\qquad &q_{k+1}=\dfrac{p_k^2-2a_4q_k^4-a_2q_k^2}{2\sqrt{a_4} q_k p_k}\,,\qquad
p_{k+1}= \dfrac{q_k ^4(2a_4q_k ^2+a_2)^2-(4a_4q_k ^4+p_k ^2)p_k^2}{4\sqrt{a_4}q_k ^2p_k ^2}\,,
\\
\\
N=3\,,\qquad &q_{k+1}=
q_k +\frac{4 q_k  p_k ^2 (2 a_4 q_k ^4+a_2  q_k ^2-p_k ^2)}{4 a_4^2 q_k ^8+4 a_2  a_4 q_k ^6+a_2 ^2 q_k ^4-8 a_4 q_k ^4 p_k ^2-2 a_2  q_k ^2 p_k ^2+p_k ^4}\,,
\\
\\
&p_{k+1}=-p_k+-\frac{(q_k-q_{k+1} ) \Bigl(a_2(q_{k+1} +q_k )+2a_4q_k ^2(3q_{k+1} -q_k )\Bigr)}{2 p_k}+\frac{ q_k ^2(q_{k} -q_{k+1} )^2 (2 a_4 q_k ^2+a_2 )^2}{2 p_k ^3}
\end{array}
\eq
\begin{prop}
Relations (\ref{M-map}) define two iterative systems of 2-point maps
\[
\cdots\,\xrightarrow[N]{}\left(
  \begin{array}{c}
    q_{k-1} \\
    p_{k-1} \\
  \end{array}\right)\xrightarrow[N]{}\left(
  \begin{array}{c}
    q_{k} \\
    p_{k} \\
  \end{array}
\right)
\xrightarrow[N]{}\left(
  \begin{array}{c}
    q_{k+1} \\
    p_{k+1} \\
  \end{array}
\right)\xrightarrow[N]{}\,\cdots
\]
which can be considered as the counterparts of usual geometric progression. These 2-points maps are canonical transformations of valence $N$ preserving the form of Hamiltonian $H$ (\ref{ell-dham}), i.e. from
$\{q_k,p_k\}=1$ and (\ref{M-map}) will follow that $\{q_{k+1},p_{k+1}\}=N$.
\end{prop}
The proof is a straightforward calculation.

Let us now take intersection divisor
\[div(X\cdot Y)=(P_1+P_2+P_3)+P_4\,,\]
see Fig.2a.  If we identify coordinates of all the intersection points $P_1,\ldots,P_4$ with the partial solutions of the Hamilton equations,
relations  (\ref{add-ell4}) and $y_4=-\mathcal P(x_4)$ define 4-point mapping
\[
\left(
  \begin{array}{c}
    q_{1},q_{2},q_3 \\
    p_{1},p_{2},p_3 \\
  \end{array}\right)
\xrightarrow[]{}\left(
  \begin{array}{c}
    q_{4} \\
    p_{4} \\
  \end{array}
\right)
\]
which has the standard properties.
\begin{prop}
Discrete map $M\times M\times M\to M$
\[
q_{4}=-q_{1}-q_{2}-q_3+\varphi(q_{1},q_2,q_3,p_1,p_2,p_3)\,,\qquad p_{4}=-(b_2q_{4}^2-b_1q_{4}-b_0)\,,
\]
where $\varphi$ and $b_k$ are given by  (\ref{add-ell4},\ref{lag-cub}), preserves the form of Hamiltonian and original Poisson bracket.
\end{prop}
The proof is a straightforward calculation.

In order to get iterative systems of finite-difference equations we identify one of the intersection points with parameter of discretization. For instance, we can substitute
\[x_1=q_{k-1},\qquad y_1=p_{k-1}\,,\qquad x_2=q_{k}\,,\qquad y_2=p_{k}\,,\qquad x_4=q_{k+1}\,,\qquad y_4=-p_{k+1}\]
and
\[
x_3=\lambda_k\,,\qquad y_3=\mu_k
\]
in (\ref{add-ell4}) in order to obtain a system of 3-point mappings
\bq\label{map-2-1}
\begin{array}{l}
q_{k+1}=-q_{k-1}-q_{k}-\lambda_k+\varphi(q_{k-1},q_k,\lambda_k,p_{k-1},p_k,\mu_k)\,\\ \\
p_{k+1}=-(b_2q_{k+1}^2-b_1q_{k+1}-b_0)\,.\\
\end{array}
\eq
Here$\varphi$ is the rational function defined  (\ref{add-ell4}, \ref{lag-cub}).

\begin{prop}
Relations (\ref{map-2-1}) determine   iterative systems of the 3-point maps
\[
\cdots\left(
  \begin{array}{c}
    q_{k-1},q_k \\
    p_{k-1},p_k \\
  \end{array}\right)
\xrightarrow[\lambda_{k}]{}\left(
  \begin{array}{c}
    q_{k+1} \\
    p_{k+1} \\
  \end{array}
\right)\,,\qquad
\left(
  \begin{array}{c}
    q_{k},q_{k+1} \\
    p_{k},p_{k+1} \\
   \end{array}
\right)
\xrightarrow[\lambda_{k+1}]{}\left(
  \begin{array}{c}
    q_{k+2} \\
    p_{k+2} \\
  \end{array}
\right)\cdots
\]
preserving the form of Hamiltonian and original Poisson bracket.
\end{prop}
The proof is a straightforward calculation in which we have to  	take into account that
 $\mu_k=\pm\sqrt{f(\lambda_k)}$ is a function on phase space, which has nontrivial Poisson brackets with   $q_1,p_1$ and $q_2,p_2$
 simultaneously, see discussion in  \cite{bob98,fed05,kuz02,sur03}.

 Let us now take intersection divisor
\[div(X\cdot Y)=(P_1+2P_2)+P_3\,,\]
see Fig. 2b.  At $a_3=a_1=0$ relation (\ref{add-21}) looks like
\bq\label{map-21}
x_3=-2x_1-x_2+\varphi\,,\qquad \varphi=\dfrac{\varphi_1}{\varphi_2}\,,
\eq
 where
 \[
 \varphi_1=2x_1\Bigl(x_1(x_1-x_2)(2a_4x_1^2+a_2) -y_1^2+y_1y_2\Bigr)
 \Bigl((x_1^2-x_2^2)(2a_4x_1^2+a_2) -2y_1^2+2y_1y_2 \Bigr)\,,
 \]
 and
 \[\begin{array}{rcl}
 \varphi_2&=&4a_4^2x_1^6(x_1-x_2)^2+\left(a_2x_1(x_1-x_2)-y_1^2+y_1y_2\right)^2\\
 \\
 &+&
 a_2(x_1-x_2)\Bigl(
 4a_2x_1^4(x_1-x_2)-(5x_1^3y_1-4x_1^3y_2-3x_1^2x_2y_1+3x_1x_2^2y_1-x_2^3y_1)y_1
 \Bigr)
 \end{array}
 \]
 Substituting  $x_i=q_i$ and $p_i=y_i$, $i=1,2,3$ in  (\ref{map-21}) and $y_3=-\mathcal P(x_3)$, one gets 3-point map which does not  Poisson, i.e. bracket $\{q_3,p_3\}$ does not function on $q_3$ and $p_3$ only\,.

 \begin{prop}
 If double  point $P_1$ plays the role of parameter
 \[
 x_1=\lambda_k\,,y_1=\pm\sqrt{p_k^2-f(q_k)+f(\lambda_k)}  \qquad x_2=q_k\,,y_2=p_k\qquad x_3=q_{k+1}\,,y_3=p_{k+1}\,,
 \]
 then relations (\ref{map-21}) and $y_3=-\mathcal P(x_3)$ define 2-point map  preserving  the form of Hamiltonian and Poisson bracket.

If ordinary point $P_2$ plays the role of the parameter
 \[
 x_1=q_k\,,y_1=p_k\,,\qquad x_2=\lambda_k\,,y_2=\pm\sqrt{p_k^2-f(q_k)+f(\lambda_k)}\qquad x_3=q_{k+1}\,,y_3=p_{k+1}\,,
 \]
 then relations (\ref{map-21}) and $y_3=-\mathcal P(x_3)$ define 2-point map  preserving  the form of Hamiltonian and Poisson bracket up to the scaling factor, i.e. from
 $\{q_k,p_k\}=1$ will follow that $\{q_{k+1},p_{k+1}\}=2$.
  \end{prop}
 The proof is a straightforward calculation.

Let us also consider the intersection of  genus one hyperelliptic curve  $X$ with line $Y$.  Substituting
$x_i=q_i$ and $y_i=p_i$ in  (\ref{line-34}) and $y_{3,4}=-\mathcal P(x_{3,4})$ one gets
\bq\label{line-map}
q_{3,4}=-\dfrac{q_1}{2}-\dfrac{q_2}{2}-\dfrac{a_3}{2a_4}\pm\dfrac{\sqrt{\alpha(q_1,q_2,p_1,p_2)}}{2a_4}\,,\qquad
p_{3,4}=-\dfrac{q_{3,4}-q_2}{q_1-q_2}p_1-\dfrac{q_{3,4}-q_1}{q_2-q_1}p_2\,,
\eq
where $\alpha(q_1,q_2,p_1,p_2)$ is given by  (\ref{alpha}).
\begin{prop}
Relations (\ref{line-map}) define invertible  algebraic 4-point mapping  $M\times M\to M\times M$
\[
\left(
  \begin{array}{c}
    q_{1},q_{2}\\
    p_{1},p_{2}\\
  \end{array}\right)
\xrightarrow[]{}\left(
  \begin{array}{c}
    q_3,q_{4} \\
    p_3,p_{4} \\
  \end{array}
\right)
\]
 preserving the form of Hamiltonian and canonical Poisson bracket.
\end{prop}
The proof is a straightforward calculation.

\section{Quintic oscillator}
\setcounter{equation}{0}
Let us consider Hamiltonian
\bq\label{g-duff}
H(q,p)=p^2-a_6q^6-a_5q^5-a_4q^4-a_3q^3-a_2q^2-a_1q\,,
\eq
and canonical Poisson bracket $\{q,p\}=1$, which  determine standard Hamilton equations
\bq\label{g-duff-eqh}
\dot{q}=\{q,H\}=2p\,,\qquad \dot{p}_j=\{p,H\}=6a_6q^5+5a_5q^4+4a_4q^3+3a_3q^2+2a_2q+a_1
\eq
and Newton equation
\bq\label{geq-duff}
\ddot{q}=12a_6q^5+10a_5q^4+8a_4q^3+6a_3q^2+4a_2q+2a_1\,.
\eq
At $a_5=a_3=a_1=0$ this system is the so-called cubic-quintic Duffing oscillator, which can be found in the modeling of free vibrations of a restrained uniform beam with intermediate lumped mass, the nonlinear dynamics of slender elastica, the generalized Pochhammer-Chree (PC) equation, the generalized compound KdV equation in nonlinear wave systems and so  on  \cite{ez13,mook73}.

 We identify a common level curve $H=E$ with the genus two hyperelliptic curve $X$ on a projective plane
\bq\label{hell-curve}
X:\qquad y^2=f(x)\,,\qquad f(x)=a_{6}x^{6}+a_{5}x^{5}+a_4x^3+a_3x_3^2+a_2x^2+a_1x+a_0\,,
\eq
where  $E=a_0$.  Integration of the equations (\ref{g-duff-eqh}) leads to the Jacobi inversion problem on the curve  $X$
\bq\label{g-duff-aj}
 \dfrac{dq}{p}=2dt\quad  \Rightarrow\quad\int^q \dfrac{dx}{\sqrt{f(x)}}=2t\,.
\eq
In \cite{dub81} we can find an impressive number the explicit solutions Jacobi inversion problems when the number
degrees of freedom $n$ is equal to a genus $g$ of hyperelliptic curve $X$.

If $n>g$ an analytic integration of the corresponding equations of motion  is possible, but it becomes more complicated, see discussion in  \cite{en03}. Thus, direct numerical integration of the equations of motion is certainly a faster way to obtain the
time course of the motion. Analytical and numerical integration of the Duffing oscillator is more easy because at $a_5=a_3=a_1=0$
Common level curve $X$ (\ref{hell-curve}) is  the so-called bielliptic curve. Nevertheless, even in this case in order to get suitable approximate solutions we have to apply the cumbersome  numerical methods:  homotopy analysis method,  homotopy Pade technique, energy balance method, combination of Newton's method and the harmonic balance method and so on, see  \cite{ez13} and references within.

 Our aim is to discuss exact discretization of  one-dimensional oscillator (\ref{g-duff-eqh},\ref{geq-duff}) associated with genus two hyperelliptic curve $X$ (\ref{hell-curve})which could be useful for exact numerical integration of the equations of motion.

\subsection{Two example of intersection divisors}
 Let us consider intersection $X$ (\ref{hell-curve}) with cubic
\[
Y:\qquad y=\mathcal P(x)\,,\qquad \mathcal P(x)=b_3x^3+b_2x^2+b_1x+b_0\,.
\]
Substituting  $y=\mathcal P(x)$ into the equation (\ref{hell-curve}) one gets Abel polynomial $\psi(x)$. The roots of this polynomial are abscissas of the intersection points which form support of the six degree intersection divisor  $div(X\cdot Y$
\[\mathrm{deg}\,div(X\cdot Y)=6,\]
 according to B\'{e}zout's theorem. In Fig.3a  we present this intersection divisor with two points at infinity
\[div(X\cdot Y)=(P_1+P_2)+P_3+P_4+2P_\infty=0\,,\]
and in the Fig.3b we present divisor with six rational ordinary points
 \[div(X\cdot Y)=(P_1+P_2+P_3+P_4)+P_5+P_6=0\,.\]
 Using brackets $(.)$  we separate a part of  the intersection divisor which is necessary for Lagrange interpolation of cubic polynomial $\mathcal P(x)$.
\begin{figure}[!ht]
\begin{minipage}[h]{0.49\linewidth}
\center{\includegraphics[width=0.85\linewidth, height=0.2\textheight]{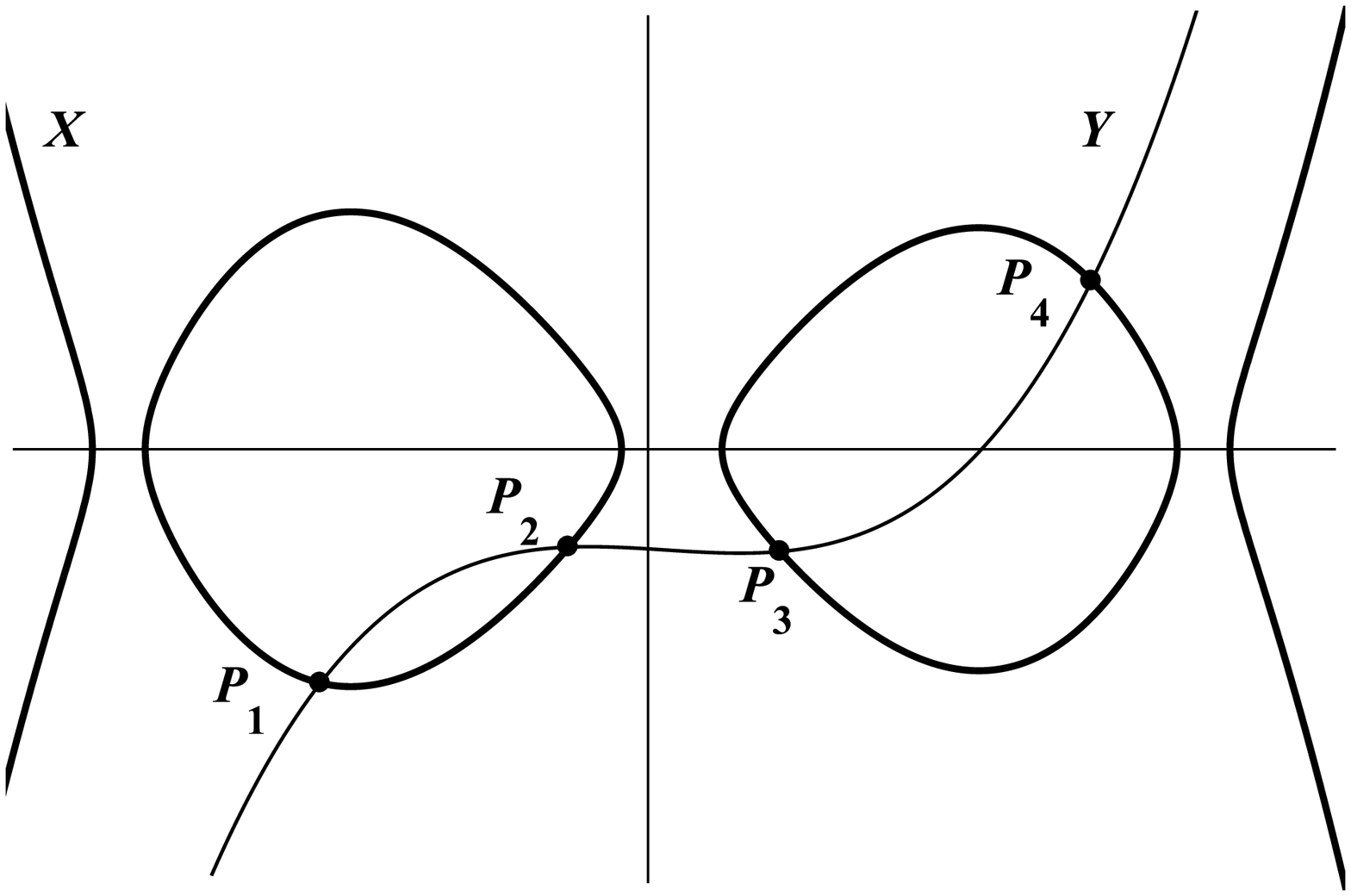} \\ a)  $(P_1+P_2)+P_3+P_4+2P_\infty=0$}
\end{minipage}
\hfill
\begin{minipage}[h]{0.49\linewidth}
\center{\includegraphics[width=0.85\linewidth,height=0.2\textheight]{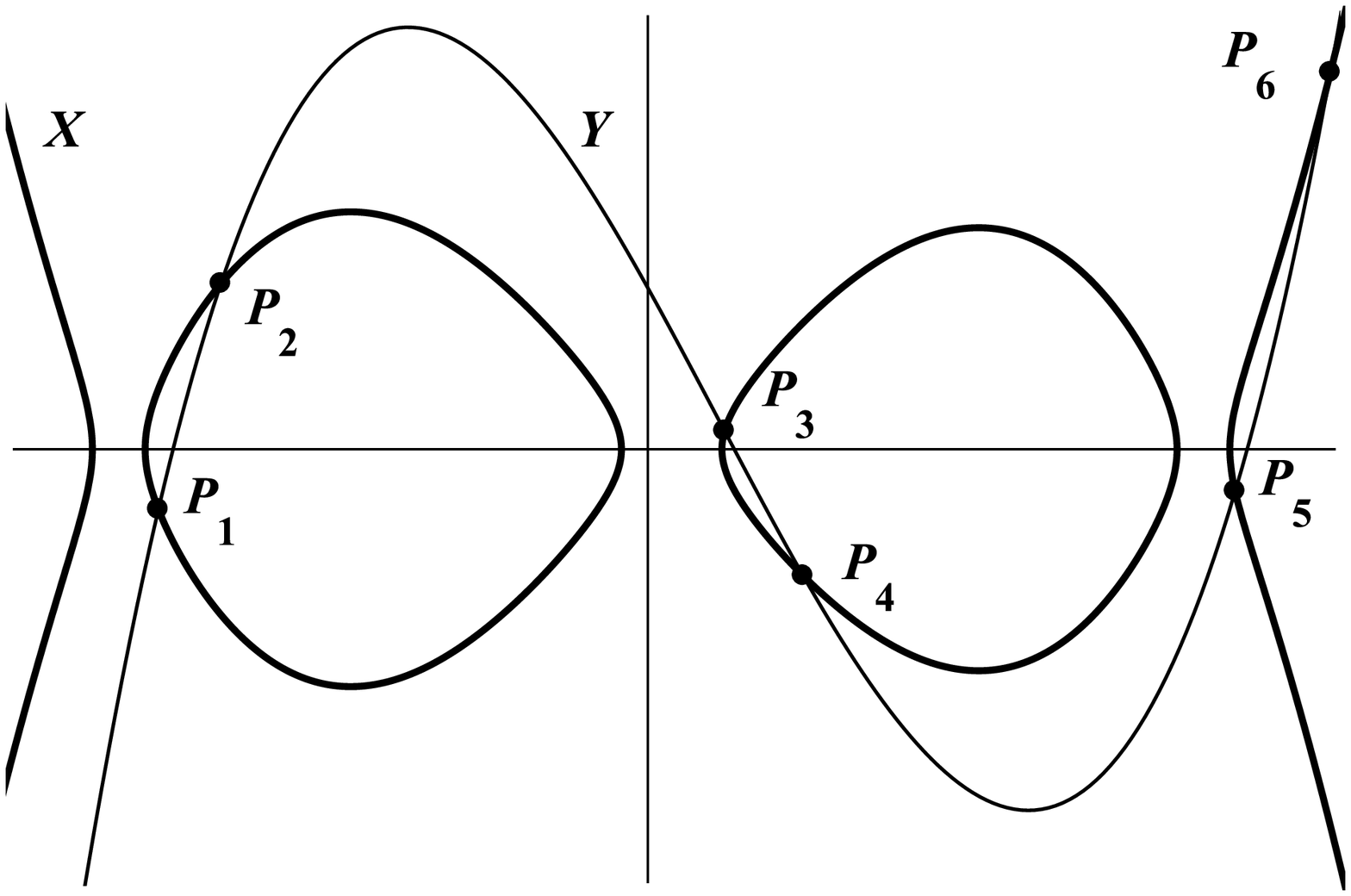} \\  b) $(P_1+P_2+P_3+P_4)+P_5+P_6=0$}
\end{minipage}
\caption{Intersection of $X$  (\ref{hell-curve}) and cubic $Y:\,y=b_3x^3+b_2x^2+b_1x+b_0$}
\end{figure}

For the intersection divisor on Fig.3a  we have
\[
a_6= b_3^2\,,\qquad\mbox{and}\qquad  a_5= 2b_2b_3\,,
\]
thus Abel's polynomial $\psi(x)=\mathcal P(x)^2-f(x)$  is equal to
\[\begin{array}{cl}
\psi&= (2 b_1 b_3 +b_2^2-a_4) x^4+(2 b_0 b_3 +2 b_1 b_2-a_3 ) x^3+(2 b_0 b_2+b_1^2-a_2) x^2+(2 b_0 b_1-a_1) x+b_0^2-a_0\\
\\
&=(2 b_1 b_3 +b_2^2-a_4)(x-x_1)(x-x_2)(x-x_3)(x-x_4).
\end{array}
\]
According  \cite{ab,bak97,gr} coefficients of this polynomial at $ x^3$ and  $x^2$ give rise to the standard equations  between abscissas of the rational intersection points
\[
\sum_{i=1}^4x_i=-\frac{a_3-2b_3b_0-2b_2b_1}{a_4-2b_3b_1-b_2^2}\,, \qquad
\sum_{i\neq j}^4 x_ix_j=\dfrac{a_2-2b_2b_0-b_1^2}{a_4-2b_3b_1-b_2^2}\,.
\]
Solving these equations with respect to  $x_3$ and $x_4$ one gets the following relations
\bq\label{hell-p4}
x_{3,4}=\sigma_{\pm}(x_1,x_2,y_1,y_2)
\eq
where
\[\begin{array}{rcl}
\sigma_\pm&=&-\dfrac{x_1}2-\dfrac{x_2}2-\dfrac{a_3-2b_0 b_3 -2b_1 b_2 }{2(a_4-2b_1 b_3 -b_2 ^2)}
\pm\dfrac12\Bigl(-3x_1^2-2x_1x_2-3x_2^2\Bigr.\\
\\
&-&\left.\dfrac{2 (-2 b_0  b_3 -2 b_1  b_2 +a_3)) (x_1+x_2)-8 b_0  b_2 -4 b_1 ^2+4 a_2}{a_4-2 b_1  b_3 -b_2 ^2}
+\dfrac{(a_3-2 b_0  b_3 -2 b_1  b_2 )^2}{(a_4-2 b_1  b_3 -b_2 ^2)^2}
\right)^{1/2}\,.
\end{array}
\]
and  $b_k$ are coefficients of the following  cubic polynomial
\bq\label{hell-pol4}
 \mathcal P(x)=
\sqrt{a_6} (x-x_1)(x-x_2)(x+x_1+x_2)+\frac{a_5}{2\sqrt{a_6}} (x-x_1)(x-x_2)+\frac{x-x_2}{x_1-x_2}y_1+\frac{x-x_1}{x_2-x_1}y_2\,.
\eq
Below we will use relations (\ref{hell-p4}) to construct various exact discretizations of our one-dimensional integrable system.

For the intersection divisor in Fig. 3b six roots $x_1,\ldots,x_6$ of the Abel polynomial
\[
\psi(x)=\mathcal P(x)^2-f(x)=(b_3^2-a_6)(x-x_1)(x-x_2)(x-x_3)(x-x_4)(x-x_5)(x-x_6)\,.
\]
satisfy to equations
\[
\sum_{i=1}^6x_i=-\frac{a_5-2b_3b_2}{a_6-b_3^2}\,, \qquad
\sum_{i\neq j}^6 x_ix_j=\frac{a_4-2b_3b_1-b_2^2}{a_6-b_3^2}\,,
\]
where $b_k$ are coefficients of the following  cubic polynomial
\bq\label{hell-pol6}
\begin{array}{rcl}
 \mathcal P(x)&=&
\dfrac{(x-x_2)(x-x_3)(x-x_4)y_1}{(x_1-x_2)(x_1-x_3)(x_1-x_4)}+\dfrac{(x-x_1)(x-x_3)(x-x_4)y_2}{(x_2-x_1)(x_2-x_3)(x_2-x_4)}\\ \\
&+&\dfrac{(x-x_1)(x-x_2)(x-x_4)y_3}{(x_3-x_1)(x_3-x_2)(x_3-x_4)}
+\dfrac{(x-x_1)(x-x_2)(x-x_3)y_4}{(x_4-x_1)(x_4-x_2)(x_4-x_3)}\,.
\\
\end{array}
\eq
Solving these equations with respect to $x_5$ and $x_6$ one gets
\bq\label{hell-p6}
x_{5,6}=\tau_{\pm}(x_1,x_2,x_3,x_4,y_1,y_2,y_3,y_4)
\eq
where
\[
\begin{array}{l}
\tau_{\pm}(x_1,x_2,x_3,x_4,y_1,y_2,y_3,y_4)=\dfrac12\left( \dfrac{2b_2  b_3 -a_5}{a_6-b_3 ^2}- \displaystyle\sum _{i=1}^4 x_i\right)\qquad\\
\\
\qquad\qquad\pm\dfrac12
\left(  -{\displaystyle \sum_{i,j=1}^4} x_ix_j  - 2{\displaystyle \sum_{i=1}^4 x_i^2}-\frac{2 (a_5-2 b_2  b_3 ) { \sum_{i=1}^4} x_i+4 a_4-8 b_3  b_1 }{a_6-b_3 ^2}
+\frac{a_5^2-4 a_5 b_2  b_3 +4 a_6 b_2 ^2}{(a_6-b_3 ^2)^2}\right)^{1/2}\,.
\end{array}
\]
These standard relations between abscissas  of the intersection points may be found in \cite{ab,bak97,gr}.  We suppose to apply these relations to construction of the  finite-difference equations  (\ref{fd-eq}) relating solutions of the equation of motion  (\ref{geq-duff}).

\subsection{Examples of finite-difference equations}
Let us consider intersection divisor  with four rational intersection points Fig.3a
\[div(X\cdot Y)=(P_1+P_2)+P_3+P_4+2P_\infty\,,\]
 Substituting solutions of the Hamilton equations (\ref{g-duff-eqh}) and parameters $\lambda_{ik}$   into the relations (\ref{hell-p4}-\ref{hell-pol4}) we can get  4-point mapping
\bq\label{4-point-m}
\left(
  \begin{array}{c}
    q_{1},q_{2} \\
    p_{1},p_{2} \\
  \end{array}\right)
\xrightarrow[]{}\left(
  \begin{array}{c}
    q_3,q_{4} \\
    q_4,p_{4} \\
  \end{array}
\right)\,,\qquad q_{3,4}=\sigma_{\pm}(q_1,q_2,p_1,p_2)\,,\qquad p_{3,4}=-\mathcal P(q_{3,4})\,,
\eq
 system of 3-point mappings depending on one parameter
\[
\left(
  \begin{array}{c}
    q_{k} \\
    p_{k}\\
   \end{array}
\right)
\xrightarrow[\lambda_{k}]{}\left(
  \begin{array}{c}
    q_{k+1},q_{k+2} \\
    p_{k+1},p_{k+2} \\
  \end{array}
\right)\,,\qquad
\begin{array}{ll}
 q_{k+1}=\sigma_{+}(q_k,\lambda_k,p_k,\mu_k)\,,\qquad &p_{k+1}=-\mathcal P(q_{k+1})\,,\\
 q_{k+2}=\sigma_{-}(q_k,\lambda_k,p_k,\mu_k)\,,\qquad &p_{k+2}=-\mathcal P(q_{k+2})\,,\\
\end{array}
\]
and system of invertible 2-point mappings depending on two parameter
\[
\left(
  \begin{array}{c}
    q_{k} \\
    p_{k} \\
  \end{array}
\right)
\xleftrightarrow[\lambda_{1k},\lambda_{2k}]{}\left(
  \begin{array}{c}
    q_{k+1} \\
    p_{k+1} \\
  \end{array}
\right)\,,\qquad
\begin{array}{ll}
 q_{k\phantom{+1}}=\sigma_{+}(\lambda_{1k},\lambda_{2k},\mu_{1k},\mu_{2k})\,,\qquad &p_{k\phantom{+1}}=-\mathcal P(q_{k})\,,\\
 q_{k+1}=\sigma_{-}(\lambda_{1k},\lambda_{2k},\mu_{1k},\mu_{2k})\,,\qquad &p_{k+1}=-\mathcal P(q_{k+1})\,.\\
\end{array}
\]
In the latter case
 \[
\mu_{ik}=\pm\sqrt{p_{k}^2+f(\lambda_{ik})-f(q_{k})}\,,
\]
when we calculate  $q_{k+1}$ and $p_{k+1}$ as functions on  $q_{k},p_{k}$  and
\[
\mu_{ik}=\pm\sqrt{p_{k+1}^2+f(\lambda_{ik})-f(q_{k+1})}\,,\qquad \lambda_{ik}\in\mathbb C\,,
\]
when we calculate $q_k$ and $p_k$ as functions on  $q_{k+1}$ and $p_{k+1}$. Here $f(x)$  is given by
 (\ref{hell-curve}).

It is easy to check that all these mappings preserve the form of discrete Hamiltonian
 \bq\label{gd-duff}
H=p_k^2-a_6q_k^6-a_5q_k^5-a_4q_k^4-a_3q_k^3-a_2q_k^2-a_1q_k\,.
\eq
Moreover, we can directly verify the following property of the first mapping.
\begin{prop}
Mapping (\ref{4-point-m}) preserves canonical Poisson bracket, i.e. from (\ref{4-point-m}) and
$\{q_1,p_1\}=1$, $\{q_2,p_2\}=1$ will follow that $\{q_{3},p_{3}\}=1$ and $\{q_{4},p_{4}\}=1$.
\end{prop}
For the mappings  depending on parameters  a direct check of the conservation of Poisson bracket was not carried out.

 If we take intersection divisor from Fig.3b with six rational intersection points
\[div(X\cdot Y)=(P_1+P_2)+P_3+P_4+P_5+P_6\,,\]
we can use relations (\ref{hell-pol6}-\ref{hell-p6}) to construct 6-point mapping
\bq\label{6-point-m}
\left(
  \begin{array}{c}
    q_{1},q_{2},q_3,q_4 \\
    p_{1},p_{2},p_3,p_4 \\
  \end{array}\right)
\xrightarrow[]{}\left(
  \begin{array}{c}
    q_5,q_{6} \\
    q_5,p_{6} \\
  \end{array}
\right)\,,\quad q_{5,6}=\tau_{\pm}(q_1,q_2,q_3,q_4,p_1,p_2,p_3,p_4)\,,\quad p_{5,6}=-\mathcal P(q_{5,6}),
\eq
with the following properties
\begin{prop}
Mappoing  (\ref{6-point-m}) preserves the form of Hamiltonian (\ref{gd-duff}) and  canonical Poisson bracket, i.e. from
$\{q_i,p_i\}=1$, $i=1,\ldots,4$ will follow that $\{q_{5},p_{5}\}=1$ and $\{q_{6},p_{6}\}=1$.
\end{prop}
The proof is a straightforward calculation by using modern computer algebra systems.

Replacing part of $x_i$ on parameters $\lambda_{ik}$ in (\ref{hell-pol6}-\ref{hell-p6}) one also gets the systems of $N$-points finite difference equations preserving the form of Hamiltonian. Among them we can separate  system of invertible 4-point  maps depending on two parameters
\[
\left(
  \begin{array}{c}
    q_{k-1},q_{k} \\
    q_{k-1},p_{k}\\
   \end{array}
\right)
\xrightarrow[\lambda_{1k},\lambda_{2k}]{}\left(
  \begin{array}{c}
    q_{k+1},q_{k+2} \\
    p_{k+1},p_{k+2} \\
  \end{array}
\right)\]
where
\[
\begin{array}{ll}
 q_{k+1}=\tau_{+}\bigl(q_{k-1},q_k,\lambda_{1k},\lambda_{2k},p_{k-1},p_k,\mu_{1k},\mu_{2k}\bigr)\,,\qquad &p_{k+1}=-\mathcal P(q_{k+1});\\ \\
 q_{k+2}=\tau_{-}\bigl(q_{k-1},q_k,\lambda_{1k},\lambda_{2k},,p_{k-1},p_k,\mu_{1k},\mu_{2k}\bigr)\,,\qquad &p_{k+2}=-\mathcal P(q_{k+2}).\\
\end{array}
\]
As above, for the mappings  depending on parameters  a direct check of the conservation of Poisson bracket was not carried out
because ordinates $\mu_{ik}$ associated with abscissas  $\lambda_{ik}$ are nontrivial functions of the phase space, see discussion of this problem for 2-point mappings in \cite{bob98,fed05,kuz02}.

\section{Conclusion}
In this paper we show how one can use the methods of the classical intersection theory to the exact discretization of the equations of motion of one-dimensional Hamiltonian systems. Similar methods are also applicable when the common level surface $X$ of first integrals can be realised as  a product of algebraic curves  using  either separation of variables or Lax representations for the given integrable system.

If we have the suitable Lax matrices, then refactorization in Poisson-Lie groups is viewed as one of the most universal mechanisms of integrability for integrable $2$-point maps \cite{bob98,d91,hiet16,kuz02,mos91,sur03}. In this note, we come back to the  Abel and Clebsch ideas in order to study $n$-point finite-difference equations sharing  integrals of motion and  Poisson bracket up to the integer scaling factor.

Another reason to conduct these calculations is related to construction of finite-difference equations  (\ref{fd-eq}) relating points on the common level surface $X$ of first integrals, which can not be realized as a product of the plane algebraic curves. In this generic case when we do not know the variables of separation or the  Lax matrices,  we can continue to study various configurations of points on algebraic surface $X$ in the framework of the standard intersection theory \cite{bak97,eh16,ful84, grif04,kl05}.

We can apply exact discretizations not only to the numerical integration of the equations of motion, but also to
\begin{itemize}
  \item construction of integrable discrete maps \cite{hiet16,kuz02,mos91,sur03, ves88,ves91};
  \item study of relations between different integrable systems \cite{ts15a,ts15b,ts15c};
  \item construction of new integrable systems \cite{ts17a,ts17b,ts17c,ts18d}.
 \end{itemize}
The main problem here is how to distinguish intersection divisors suitable  for these purposes.

The work was supported by the Russian Science Foundation  (project  18-11-00032).

\end{document}